\documentclass[journal]{IEEEtran}
\usepackage{graphicx}
\usepackage{etoolbox}
\usepackage{multirow}
\usepackage{array}
\usepackage{diagbox}
\usepackage{mathtools}
\usepackage{amsmath}
\usepackage{amssymb}
\usepackage{empheq}
\usepackage{bm}
\usepackage{color}
\usepackage{amsthm}
\usepackage{subcaption}
\usepackage{comment}
\usepackage{algorithm}
\usepackage{algorithmicx}
\usepackage{algpseudocode}

\allowdisplaybreaks
\IEEEoverridecommandlockouts

\def\BibTeX{{\rm B\kern-.05em{\sc i\kern-.025em b}\kern-.08em
    T\kern-.1667em\lower.7ex\hbox{E}\kern-.125emX}}

\IEEEaftertitletext{\vspace{-2\baselineskip}}

\begin{document}
\title{Soft Robotics-Inspired Flexible Antenna Arrays}
\author{\author{Elio Faddoul, \IEEEmembership{Member, IEEE}, Andreas Nicolaides, \IEEEmembership{Member, IEEE},\\ Konstantinos Ntougias, \IEEEmembership{Member, IEEE}, and Ioannis Krikidis, \IEEEmembership{Fellow, IEEE}
\thanks{The authors are with the IRIDA Research Centre for Communication Technologies, Department of Electrical and Computer Engineering, University of Cyprus, Cyprus (email: \{efaddo01, anicol09, kntoug01, krikidis\}@ucy.ac.cy).}}}

\maketitle

\begin{abstract}
In this work, a novel soft continuum robot-inspired antenna array is proposed, featuring tentacle-like structures with multiple antenna elements. The proposed array achieves reconfigurability through continuous deformation of its geometry, in contrast to reconfigurable antennas which incur a per-element control. More specifically, the deformation is modeled by amplitude and spatial frequency parameters. We consider a multi-user multiple-input single-output downlink system, whereby the optimal deformation parameters are found to maximize the sum rate in the network. A successive convex approximation method is adopted to solve the problem. Numerical results show that the proposed deformable array significantly outperforms fixed geometry and per-element reconfigurable arrays in sum rate, demonstrating the benefits of structure-level flexibility for next-generation antenna arrays.
\end{abstract}


\section{Introduction}
Antenna and radio frequency (RF) technologies have significantly advanced in recent years to address the evolving demands of next-generation wireless communication systems~\cite{6G}. In particular, multiple-input multiple-output (MIMO) systems have gained substantial attention due to their diversity and multiplexing gains. However, traditional antenna arrays rely on fixed metallic structures, limiting their performance in meeting the demands of emerging applications. This challenge has motivated the exploration of reconfigurable antenna (RA) solutions through metamaterials, such as reconfigurable intelligent surfaces~\cite{RIS}, fluid antennas (FAs)~\cite{FA}, movable antennas (MAs)~\cite{MA}, flexible antenna arrays (FAAs)~\cite{FlAA}, and flexible intelligent metasurfaces (FIMs)~\cite{FIM}.

Generally, RA arrays offer the capability of dynamically repositioning their antenna elements within a specified region to unlock additional spatial degrees of freedom (DoFs) with a limited number of physical antennas~\cite{FAtut,MAtut}. For instance, the work in~\cite{FAMIMO} considers the joint optimization of transmit/receive FA positions to maximize the achievable rate in single-user MIMO systems, whereas transmit sum-power minimization is studied in~\cite{PowConMA} for multi-user multiple-input single-output (MISO) downlink systems with an MA-equipped base station (BS). Typically, the realization of such arrays includes per-element control based on mechanical designs with stepper motor-driven metallic structures for MAs, or liquid-metal designs controlled by micro-pumps for FAs, which could incur high cost and power consumption~\cite{MAtut},~\cite{PowConMA}. In addition, the work in~\cite{FIMSensing} jointly optimizes the transmitting FIM’s 3D surface shape to maximize the cumulated probing power at multiple target locations for wireless sensing applications.

More recently, soft continuum robots are emerging as an innovative class of robotic systems inspired by biological structures such as octopus tentacles, offering enhanced flexibility compared to traditional rigid-bodied robots~\cite{SRnature}. Recent advancements have demonstrated the effective integration of soft robots into RA systems, enabling dynamic tuning of electromagnetic characteristics, such as beamforming and radiation pattern reconfiguration~\cite{TipSR}. For instance, antennas based on soft continuum robots provide continuous and real-time adjustments of antenna shapes and electromagnetic responses, demonstrating their potential in wireless communication scenarios~\cite{RotSR}. This technology, supported by fabrication methods like 3D printing and additive manufacturing, offers a technologically mature solution for overcoming the existing limitations of fixed antenna and RA systems. So far, a mathematical framework based on such structures has not been provided.

Inspired by recent advances in soft continuum robotics and the natural movement of octopus tentacles, we introduce a novel soft robot antenna (SRA) design. Specifically, we focus on a MISO downlink communication system, where the BS is equipped with an SRA array that consists of multiple tentacle-like structures, each comprising several antenna elements. Flexibility is achieved by dynamically changing the geometric structure of the array. We note that in contrast to the existing FA, MA, and FIM architectures, the proposed antenna array achieves flexibility through structure deformation rather than repositioning the individual elements. To the authors' best knowledge, this is the first work that investigates the potential performance benefits of a soft continuum robot-based antenna array in wireless communications. To this end, we model the deformation of the SRA array through controllable parameters, such as amplitude and spatial frequency, thus capturing realistic physical movements. These parameters are optimized to maximize the sum rate under a transmit sum-power constraint, assuming a heuristic linear precoder. To handle this non-convex problem, we resort to a successive convex approximation (SCA) technique to iteratively solve for the optimal deformation parameters. Numerical results reveal that the proposed soft robotic-inspired antenna array provides substantial performance gains in terms of sum rate compared to the fixed position counterpart (up to $73\%$) as well as the per-element RA arrays (up to $26\%$), emphasizing the importance of structure flexibility for next-generation antenna arrays.

\textbf{Notation:} Lower and upper case boldface letters denote vectors and matrices, respectively; $[\cdot]^T, [\cdot]^H$ and $\rm{Tr}(\cdot)$ are the transpose, Hermitian, and trace operators, respectively; $\odot$ is the Hadamard product; $|\cdot|$ and $\lVert \cdot \rVert$ refer to the absolute value of a scalar and the $\ell_2$ norm of a vector, respectively; $\mathcal{CN}(\mu,\sigma^2)$ denotes a complex Gaussian distribution with mean $\mu$ and variance $\sigma^2$; $\mathbb{E}[\cdot]$ represents the expectation operator and $\mathbf{I}_N$ is the $N\times N$ identity matrix.
\begin{figure*}
\centering
\begin{subfigure}[t]{0.32\textwidth}
  \centering
  \includegraphics[width=0.78\columnwidth]{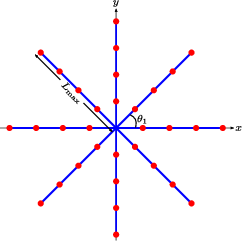}
\caption{No deformation.}\label{fig:noDeformation}
\end{subfigure}%
~
\begin{subfigure}[t]{0.32\textwidth}
  \centering
  \includegraphics[width=\columnwidth]{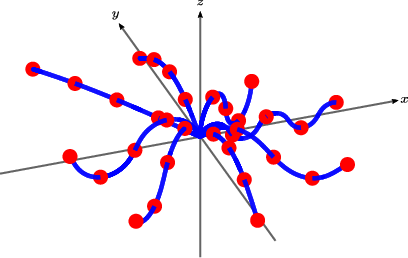}
  \caption{Random deformation in 3D space.}\label{fig:randomDeformation3D}
\end{subfigure}
~
\begin{subfigure}[t]{0.32\textwidth}
  \centering
  \includegraphics[width=0.78\columnwidth]{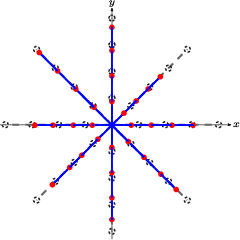}
  \caption{Projected deformation in 2D space.}\label{fig:randomDeformation2D}
\end{subfigure}
\vspace{-1mm}
\caption{An abstraction of an SRA array at an arbitrary time $t$, with $M=8$ tentacles, each having $N=4$ antenna elements. An antenna array is represented by a blue segment, while an antenna element is depicted by a red circle.}
\vspace{-2mm}
\label{fig:1}
\end{figure*}

\section{System Model}
We consider a multi-user MISO downlink system between a BS and $K$ single fixed position antenna user terminals. The BS is equipped with a soft robot-inspired FAA consisting of $M$ arms, termed tentacles, each supporting $N$ antenna elements, as depicted in Fig. \ref{fig:1}. Due to tentacle motion, the antenna elements dynamically change their position, thereby altering the radiation pattern~\cite{RotSR}.
\vspace{-2mm}
\subsection{Tentacle-Based Antenna Array Model}
We consider the representation of the $m$-th tentacle, $m\in \mathcal{M}=\left\{1,\ldots,M\right\}$, as a continuous space curve of fixed total arc length $L_{\rm{max}}$. We denote by $\ell \in \left[0, L_{\rm{max}}\right]$ an arbitrary point within a tentacle, and orient the $m$-th tentacle in the $xy$-plane at azimuth $\theta_{m} = \frac{2\pi m}{M}$, as shown in Fig. \ref{fig:noDeformation}. The deformation of each tentacle is assumed to follow a sinusoidal pattern since it is mathematically tractable and captures the basic deformation behaviour of soft continuum structures~\cite{TipSR}. Under such dynamic deformation, we superpose a vertical displacement, in which the 3D position of the point $\ell$ at the $m$-th tentacle is given by\footnote{The soft robotic antenna can be realized by an inverse pneumatically actuated muscle, which alters its shape by increasing/decreasing the pressure within the rubber tube, allowing for a low power consumption compared to stepper motor-driven MAs~\cite{PowConMA},~\cite{RotSR}.}
\begin{equation}
    \mathbf{r}_m\left(\ell, t\right) =
    \begin{bmatrix}
		x_{m}\left(\ell,t\right) \\[1mm]
		y_{m}\left(\ell,t\right) \\[1mm]
		z_{m}\left(\ell,t\right)
    \end{bmatrix}=
    \begin{bmatrix}
    	u_m\left(\ell, t\right) \cos\left(\theta_m\right) \\[1mm]
        u_m\left(\ell, t\right) \sin\left(\theta_m\right) \\[1mm]
        A_m \sin\left(\omega t + v_m \ell\right)
    \end{bmatrix},
\vspace{-1mm}
\end{equation}
where $z_{m}\left(\ell,t\right)$ is the vertical sinusoid, and $u_{m}\left(\ell, t\right)$ is its projection onto the $xy$-plane. Moreover, the angular frequency $\omega$ and time $t$ parameters are introduced to represent the dynamic oscillation of the tentacle. Note that in the case of no deformation, the $m$-th tentacle lies in the $xy$-plane so that its horizontal projection satisfies $u_{m}(\ell,t) = \ell$ at time $t$. The amplitude of the deformation is denoted by $A_m \in \left[0, A_{\rm{max}}\right]$, and $v_m \in \left[0, v_{\rm{max}}\right]$ is the spatial frequency.

Throughout this work, we assume that the array flexibility occurs along the $z$-axis, i.e., $A_m$ and $v_m$ are responsible for the movement of the antenna elements which affects the radiation pattern. Subsequently, each tentacle undergoes a retraction that is visible on the $xy$-plane, while having the overall arc length fixed, as illustrated in Figs. \ref{fig:randomDeformation3D} and \ref{fig:randomDeformation2D}. Moreover, the deformed array is depicted on top of the original geometry, i.e., without deformation, which is indicated by gray dashed lines with dashed circles (refer to Fig. \ref{fig:randomDeformation2D}). Thus, for a given amplitude and spatial frequency, the length $u_{m}\left(\ell, t\right)$ must be computed. We assume that the SRA geometry is optimized once per channel coherence interval and then held fixed during that block. Thus, we drop the functional dependencies on the time $t$. To ensure that $\ell$ remains the true arc length parameter, i.e., the total curve length is exactly $L_{\rm{max}}$ even after deformation, the tangent vector must satisfy
\begin{align}
\left \lVert \frac{\partial \mathbf{r}_m}{\partial \ell} \right \rVert &=\sqrt{ \left(\frac{\partial u_m}{\partial \ell} \right)^2 + \left(\frac{\partial z_m}{\partial \ell}\right)^2} = 1.
\end{align}
Since $\frac{\partial z_m}{\partial \ell} = A_m v_m \cos(v_m \ell)$, solving for $\frac{\partial u_m}{\partial \ell}$, we obtain
\begin{align}\label{eq:partialDerivative}
\frac{\partial u_m}{\partial \ell} = \sqrt{1-\left( A_m v_m \cos(v_m \ell) \right)^2}.
\end{align}
Ultimately, the projected tentacle length onto the $xy$-plane is
\begin{align}\label{eq:projLength}
u_m\left(\ell\right) = \int_0^{\ell} \sqrt{1 - \left[A_m v_m \cos(v_m \bar{\ell})\right]^2}\rm{d}\bar{\ell},
\end{align}
which reduces to $u_m\left(\ell\right) = \ell$ when $A_m=0$ and satisfies $u_m\left(\ell\right)<\ell$ for any nonzero $A_m$, thus validating that vertical deformation contracts the planar footprint of the tentacle without changing its total length. Note that in \eqref{eq:projLength} the condition $\left|A_m v_m\right| \leq 1$ needs to be satisfied to guarantee that the integrand remains real.

Based on the above representation, the positions of the discrete antenna elements of the array can be obtained. As previously mentioned, each tentacle contains $N$ discrete antenna elements, where the position of the $n$-th antenna along the $m$-th tentacle is $\mathbf{r}_{m,n} = \mathbf{r}_m(\ell_n)$, $n\in \mathcal{N}=\left\{1,\ldots,N\right\}$. Elements are uniformly distributed along the arc length, i.e., $\ell_n = \frac{n}{N}L_{\rm{max}}$. With $L_{\rm{max}}=N\lambda/2$, the curve spacing is $\Delta\ell = \lambda/2$. Since the deformation preserves arc length, this spacing is maintained, thus mitigating mutual coupling. Therefore, the position of the $n$-th antenna element on the $m$-th tentacle is written as
\begin{equation}
\mathbf{r}_{m,n} = 
\begin{bmatrix}
	x_{m,n} \\[1mm]
	y_{m,n} \\[1mm]
	z_{m,n}
\end{bmatrix} =
\begin{bmatrix}
	u_m\left(\ell_{n}\right) \cos\left(\theta_m\right) \\[1mm]
	u_m\left(\ell_{n}\right) \sin\left(\theta_m\right) \\[1mm]
	A_m \sin\left(v_m\ell_{n}\right)
\end{bmatrix}.
\end{equation}
Note that $u_m(\ell)$ depends on ($A_m, v_m$) through the arc length relation in \eqref{eq:partialDerivative}. Thus, $x_{m,n}$ and $y_{m,n}$ change implicitly with ($A_m, v_m$). Although the deformation is parametrized via $z_{m,n}$, the resulting element positions are updated in all three coordinates. Based on the proposed tentacle-based design for the SRA, we move on to describe the adopted channel model.

\subsection{Channel and Signal Model}
We define $\mathbf{z}_m \triangleq \left[z_{m,1}, \ldots, z_{m,N}\right]^T \in \mathbb{R}^{N\times 1}$ as the position vector of all the antenna elements of the $m$-th tentacle along the $z$-axis. Correspondingly, $\mathbf{z} = \left[\mathbf{z}_1^T, \ldots, \mathbf{z}_M^T\right]^T \in \mathbb{R}^{MN \times 1}$ represents the positions of all the antenna elements of the antenna array along the $z$-axis. According to the geometric Saleh-Valenzuela channel model, the channel $\mathbf{h}_{k}\left(\mathbf{z}\right) \in \mathbb{C}^{MN\times 1}$ from all the $MN$ elements of the SRA array to the $k$-th user is given by
\begin{align}
\mathbf{h}_{k}\left(\mathbf{z}\right)=& \sqrt{\frac{MN}{N_c N_p}} \sum_{c\in\mathcal{N}_c} \sum_{p\in\mathcal{N}_p} \beta_{c, p, k} \nonumber \\
& \qquad \quad \times \mathbf{a}\left(\vartheta_{c, p, k}, \varphi_{c, p, k}\right) \odot \mathbf{e}\left(\vartheta_{c, p, k}, \varphi_{c, p, k}\right),
\end{align}
where $N_c$ and $N_p$ are the number of scattering clusters and paths per cluster, respectively, and $\beta_{c, p, k} \sim \mathcal{CN}(0,1)$ denotes the complex gain of the $p$-th path in the $c$-th cluster, $p \in \mathcal{N}_p \triangleq \left\{ 1,\ldots, N_p \right\}$, $c \in \mathcal{N}_c \triangleq \left\{ 1,\ldots, N_c \right\}$. Furthermore, the steering vector $\mathbf{a}\left(\vartheta_{c, p, k}, \varphi_{c, p, k}\right) \in \mathbb{C}^{MN\times 1}$ is a column-stacked vector, i.e., $\mathbf{a} \triangleq \left[\mathbf{a}_{1}^{T}, \ldots, \mathbf{a}_{M}^{T}\right]^T$, in which the $n$-th entry of the $m$-th column is given by
\begin{equation}
\left[\mathbf{a}_m\left(\vartheta, \varphi\right)\right]_n = e^{-\frac{j2\pi}{\lambda}\left(x_{m,n}\sin\vartheta \cos \varphi + y_{m,n}\sin\vartheta \sin \varphi + z_{m,n} \cos\vartheta\right)},
\end{equation}
where $\vartheta_{c, p, k} \in [0, \pi/2]$ and $\varphi_{c, p, k} \in [0, 2\pi]$ correspond to the elevation and azimuth angles, respectively, and $\lambda$ is the transmission wavelength. In this work, we consider a directional antenna radiation pattern represented by $\mathbf{e}\left(\vartheta_{c, p, k}, \varphi_{c, p, k}\right) \in \mathbb{C}^{MN \times 1}$. For omnidirectional antennas, this term is set to unity. For a directional antenna oriented along the $z$-axis, the cosine pattern is considered and given by~\cite{FlAA},~\cite[Ch. 15]{BalAnt}
\begin{equation}
Q_E(\vartheta, \varphi)=
\begin{cases}
Q \cos ^\kappa \vartheta, & \vartheta \in\left[0, \frac{\pi}{2}\right], \varphi \in[0,2 \pi] \\
0, & \text { otherwise }
\end{cases},
\end{equation}
where $Q=2(\kappa + 1)$ is a normalization factor, and $\kappa \geq 1$ represents the directivity sharpness factor. Ultimately, by assuming that $\mathbf{e} \triangleq \left[\mathbf{e}_{1}^{T}, \ldots, \mathbf{e}_{M}^{T}\right]^T$, the $n$-th entry of the $m$-th column is written as
\begin{equation}
\left[\mathbf{e}_m\left(\vartheta_{c, p, k}, \varphi_{c, p, k}\right)\right]_n = \sqrt{Q_E\left(\vartheta_{c, p, k}, \varphi_{c, p, k}\right)}.
\end{equation}

We can express the channel from all $MN$ elements to all $K$ users as $\mathbf{H}\left(\mathbf{z}\right) = \left[ \mathbf{h}_{1}\left(\mathbf{z}\right), \ldots, \mathbf{h}_{K}\left(\mathbf{z}\right) \right]^T \in \mathbb{C}^{K\times MN}$. Under the assumption of linear precoding, the transmitted signal $\mathbf{x}\left(\mathbf{H}\right) \in \mathbb{C}^{MN\times 1}$ is given by
\begin{equation}
\mathbf{x}\left(\mathbf{H}\right) = \sum_{k\in \mathcal{K}} \mathbf{w}_k(\mathbf{h}_k) s_k = \mathbf{W}(\mathbf{H}) \ \mathbf{s},
\end{equation}
where $\mathbf{w}_k(\mathbf{h}_k) \in \mathbb{C}^{MN\times 1}$ denotes the precoding vector and $s_k \sim \mathcal{CN}\left(0,1\right)$ represents the independent and identically distributed (i.i.d) symbol of the $k$th user, $k \in \mathcal{K}\triangleq \left\{1,\ldots, K\right\}$. Moreover, $\mathbf{W}(\mathbf{H}) = \left[\mathbf{w}_1(\mathbf{h}_1), \ldots, \mathbf{w}_K(\mathbf{h}_K)\right] \in \mathbb{C}^{MN\times K}$ refers to the precoding matrix, and $\mathbf{s} = \left[s_1, \ldots, s_K\right]^T \in \mathbb{C}^{K \times 1}$ stands for the i.i.d transmitted symbols vector, with $\mathbb{E}\left[\mathbf{s}\mathbf{s}^H\right] = \mathbf{I}_K$. For a transmit power budget $P_{\rm{max}} \geq 0$, the transmit sum-power $P_t = \mathbb{E}\left[\left\|\mathbf{x}\left(\mathbf{H}\right)\right\|^2\right]$ is constrained as
\vspace{-2mm}
\begin{equation}
P_t =  \sum_{k\in\mathcal{K}}\left\|\mathbf{w}_k\left(\mathbf{h}_k\right)\right\|^2 = \operatorname{Tr}\left(\mathbf{W}^{H}\left(\mathbf{H}\right)\mathbf{W}\left(\mathbf{H}\right)\right)\leq P_{\rm{max}}.
\end{equation}

We consider the heuristic zero-forcing (ZF) linear precoder\footnote{ZF is employed as a standard baseline. More sophisticated precoding schemes as well as other network topologies are left for future consideration.}, which is defined as
\begin{equation}
\mathbf{W} = \alpha\mathbf{F} = \alpha\left[\mathbf{f}_1,\dots,\mathbf{f}_K\right], \ \alpha = \sqrt{\frac{P_{\max}}{\operatorname{Tr}\left(\mathbf{F}\mathbf{F}^{H}\right)}},
\end{equation}
where the ZF precoding matrix is expressed as
\vspace{-1mm}
\begin{equation}
\mathbf{F} = \mathbf{H}\left(\mathbf{H}^{H}\mathbf{H}\right)^{-1},
\vspace{-1mm}
\end{equation}
and $\mathbf{F}\in\mathbb{C}^{M\times K}$ is the non-normalized precoding matrix. $\mathbf{f}_k\in\mathbb{C}^{M}$ refers to the non-normalized precoding vector for the $k$-th user, and $\alpha$ denotes a normalization factor which ensures that the transmit sum-power constraint is satisfied. Ultimately, the received signal at the $k$th user is given by
\begin{align}
y_k &= \mathbf{H}^{H}\left(\mathbf{z}\right)\mathbf{W}\mathbf{s} + \eta_k \nonumber \\
&= \mathbf{h}_k^{H}\left(\mathbf{z}\right)\mathbf{w}_k s_k + \mathbf{h}_k^{H}\left(\mathbf{z}\right) \sum_{i\in\mathcal{K}\setminus{k}}\mathbf{w}_i s_i + \eta_k,
\end{align}
where $\eta_{k} \sim \mathcal{CN}(0,\sigma^2)$ represents the additive white Gaussian noise at user $k$.

\subsection{Problem Formulation}
Based on the above, the signal-to-interference-plus-noise ratio (SINR) of the $k$-th user is given by
\begin{align}\label{eq:SINR0}
\gamma_{k}\left(\mathbf{z}\right) = \frac{\left|\mathbf{h}_k^{H}\left(\mathbf{z}\right)\mathbf{w}_k\left(\mathbf{h}_k\right)\right|^2}{\sum\limits_{i\in\mathcal{K}\setminus\{k\}}\left|\mathbf{h}_k^H\left(\mathbf{z}\right)\mathbf{w}_i\left(\mathbf{h}_k\right)\right|^2 + \sigma^2}.
\end{align}
Thus, under Gaussian signaling, the achievable rate of the $k$-th user and the sum-rate are given by $R_k\left(\mathbf{z}\right)=\log_2\left(1+\gamma_k\left(\mathbf{z}\right)\right)$ and $R\left(\mathbf{z}\right)=\sum_{k\in\mathcal{K}}R_k\left(\mathbf{z}\right)$, respectively.

Given the ZF precoder $\mathbf{F}$, we optimize the deformation amplitudes $A_m$ and spatial frequencies $v_m$, for all $m\in \mathcal{M}$, to maximize the sum rate. This problem is formulated as
\vspace{-1mm}
\begin{subequations}
\begin{alignat}{2}
&&&\text{(P1): }\underset{\{A_m,v_m\}_{m\in\mathcal{M}}}{\max} \sum_{k\in\mathcal{K}}\!\log_2\!\left(\!\!1\!+\!\frac{\left|\mathbf{h}_k^H\left(\mathbf{z}\right)\mathbf{w}_k\left(\mathbf{h}_k\right)\right|^2}
{\!\!\!\sum\limits_{i\in\mathcal{K}\setminus\{k\}}\!\!\!\left|\mathbf{h}_k^H\left(\mathbf{z}\right)\mathbf{w}_i\left(\mathbf{h}_k\right)\right|^2 \!\!+\! \sigma^2}\!\!\!\right)\label{eq:P1a} \\
&&&\text{s.t.} \ \ \ \ 0 \leq A_m \leq A_{\rm{max}}, \quad \forall m\in\mathcal{M},\label{eq:P1b} \\
&&& \ \ \ \ \ \ \ 0 \leq v_m \leq v_{\rm{max}}, \quad \forall m\in\mathcal{M},\label{eq:P1c} \\
&&& \ \ \ \ \ \ \ \left| A_m v_m \right| \leq 1, \quad \forall m\in\mathcal{M},\label{eq:P1d}
\end{alignat}
\end{subequations}
where $z_{m,n} = A_m \sin\left(\omega t + v_m\ell_{n}\right)$. Problem (P1) is non-convex since the objective function is a sum of log of ratio terms that nonlinearly depend on the sinusoidal deformation parameters of the SRA array. In the next section, we leverage the SCA technique to iteratively solve for the optimal deformation parameters that maximize the network sum rate.


\begin{figure*}
\centering
\begin{subfigure}[t]{0.32\textwidth}
  \centering
  \includegraphics[width=\linewidth]{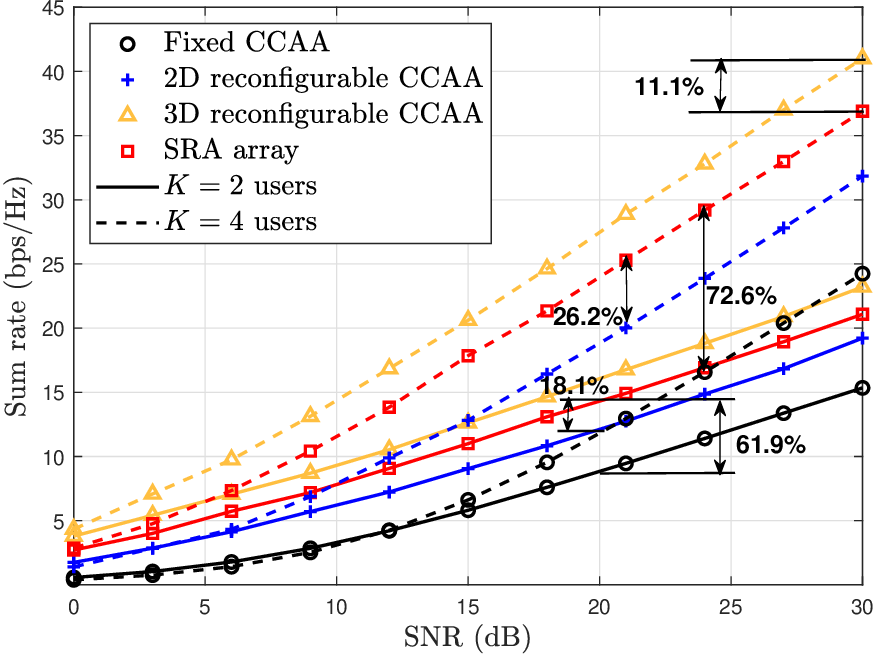}
	\caption{}\label{fig:2a}
\end{subfigure}%
~
\begin{subfigure}[t]{0.32\textwidth}
  \centering
  \includegraphics[width=\linewidth]{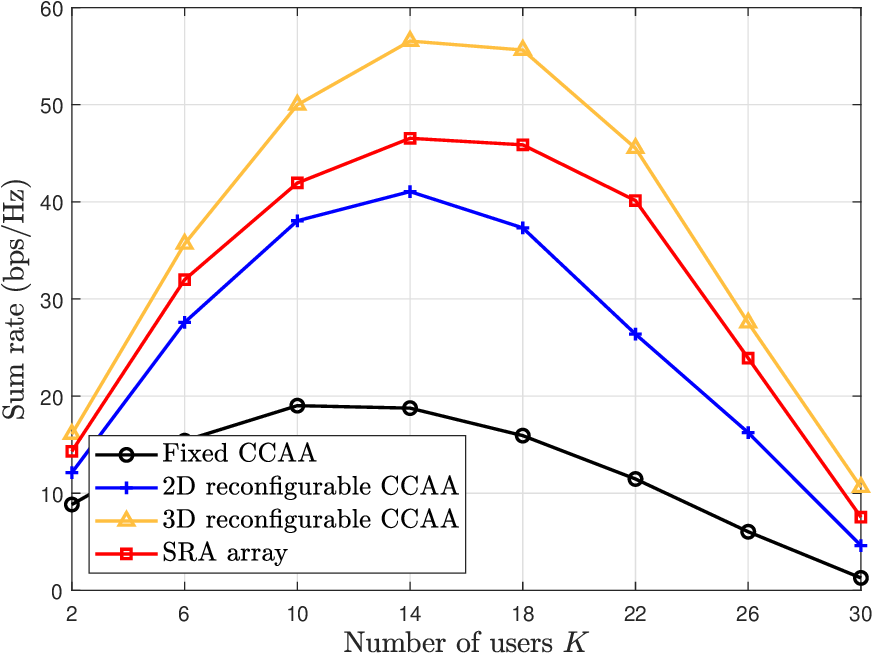}
	\caption{}\label{fig:2b}
\end{subfigure}
~
\begin{subfigure}[t]{0.32\textwidth}
  \centering
	\includegraphics[width=\linewidth]{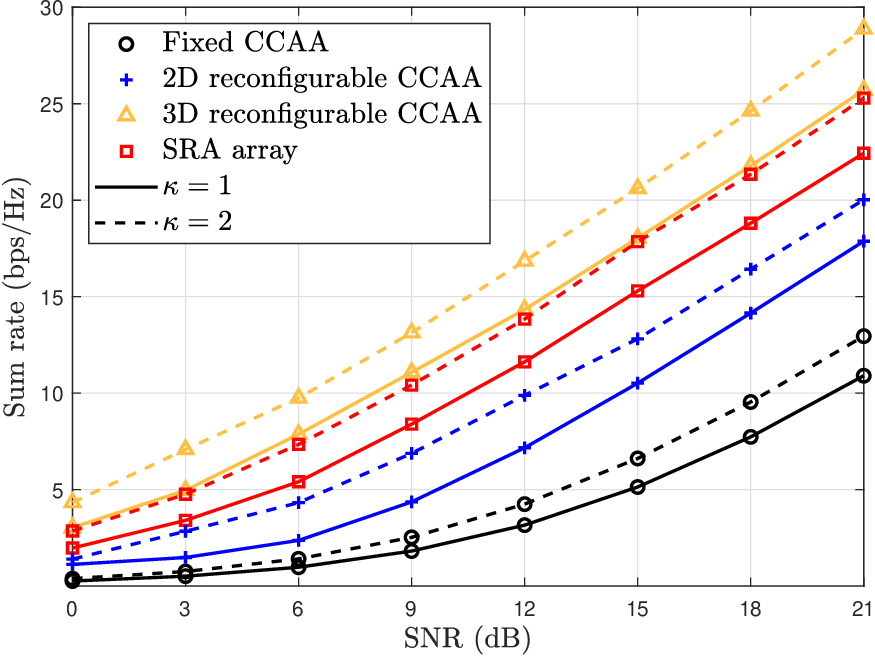}
	\caption{}\label{fig:2c}
\end{subfigure}
\vspace{-1.2mm}
\caption{(a) Sum rate vs. SNR. (b) Sum rate vs. number of users. (c) Sum rate vs. SNR for different directivity factors.}
\vspace{-2mm}
\end{figure*}


\section{Proposed Solution}
\vspace{-1.5mm}
To tackle this problem, we remark first that since the deformation occurs along each tentacle rather than individual antenna element movement, we apply a dimensionality reduction to reduce the search space. More specifically, we introduce the vector $\bar{\mathbf{z}} \in \mathbb{R}^{2M\times 1} = [A_1, \ldots, A_M, v_1, \ldots, v_M]^{T}$, and define a mapping function $g\!\!:$ $\mathbb{R}^{2M\times 1} \rightarrow \mathbb{R}^{MN\times 1}$. Thus, $\mathbf{z} = g(\bar{\mathbf{z}})$ where $\left[g(\bar{\mathbf{z}}) \right]_{(m-1)N+n} = A_m \sin(v_m \ell_n)$. The objective function then becomes
\vspace{-2mm}
\begin{align}
R(\bar{\mathbf{z}}) = \sum_{k\in\mathcal{K}} \log_2(1+\gamma_{k}(\bar{\mathbf{z}})),
\vspace{-2mm}
\end{align}
where $\gamma_{k}(\bar{\mathbf{z}})$ is obtained by replacing $\mathbf{z}$ by $\bar{\mathbf{z}}$ in \eqref{eq:SINR0}. By invoking a first-order Taylor expansion, $R(\bar{\mathbf{z}})$ is expressed as
\begin{align}\label{TaylorR}
R(\bar{\mathbf{z}}) \approx R(\bar{\mathbf{z}}^{(\rm{i})}) + \nabla_{\bar{\mathbf{z}}} R(\bar{\mathbf{z}}^{(\rm{i})})^T (\bar{\mathbf{z}} - \bar{\mathbf{z}}^{(\rm{i})}),
\vspace{-2mm}
\end{align}
where $\nabla_{\bar{\mathbf{z}}} R(\bar{\mathbf{z}}^{(\rm{i})}) \in \mathbb{R}^{2M \times 1}$ is the gradient vector over $\bar{\mathbf{z}}$ and $\rm{i}$ refers to the ${\rm{i}}$-th iterative step. The gradient calculation in \eqref{TaylorR} is given in the Appendix. Thus, (P1) is reformulated as
\vspace{-1mm}
\begin{subequations}
\begin{alignat}{2}
&&&\text{(P2): }\underset{\{A_m,v_m\}_{m\in\mathcal{M}}}{\max} R(\bar{\mathbf{z}}^{(\rm{i})}) + \nabla_{\bar{\mathbf{z}}} R(\bar{\mathbf{z}}^{(\rm{i})})^T (\bar{\mathbf{z}} - \bar{\mathbf{z}}^{(\rm{i})}) \label{eq:P2a} \\
&&&\text{s.t.} \ \ \ \ 0 \leq A_m \leq A_{\rm{max}}, \quad \forall m\in\mathcal{M},\label{eq:P2b} \\
&&& \ \ \ \ \ \ \ 0 \leq v_m \leq v_{\rm{max}}, \quad \forall m\in\mathcal{M}, \\
&&& \ \ \ \ \ \ \ \left| A_m v_m^{(\rm{i})} + A_m^{(\rm{i})} v_m - A_m^{(\rm{i})} v_m^{(\rm{i})} \right| \leq 1, \ \forall m\in\mathcal{M}.\label{eq:P2d}
\end{alignat}
\end{subequations}
Note that the constraint in \eqref{eq:P2d} is obtained by also applying a first-order Taylor expansion, to which the steps are omitted here due to space limitation. Finally, at each iteration, problem (P2) can be efficiently solved by standard convex solvers such as CVX, and the algorithm converges to the optimal solution until the difference between consecutive sum rate values is small enough or the maximum number of iterations is reached The overall procedure of the algorithm is outlined in Algorithm 1. The algorithm exhibits a complexity of $\mathcal{O}\left( N_c N_p K M N + K^2 M N \right)$ for the sum rate and gradient calculations, whereas the complexity of solving (P2) via CVX is $\mathcal{O}\left( M^{3.5}\right)$ (assuming the use of an interior-point method). Thus, for $I_{\rm{max}}$ iterations, the overall algorithm complexity is $\mathcal{O}\left( I_{\rm{max}}\left(N_c N_p K M N + K^2 M N + M^{3.5}\right)\right)$. We note that SCA is sufficient to introduce the new SRA paradigm. Other mathematical frameworks could be considered for future work.

\begin{algorithm}[!t]
\centering
\small
\begin{algorithmic}
  \State Initialize number of tentacles $M$, elements per tentacle $N$, $K$, tolerance $\varepsilon$, iteration index $\rm{i}$, max iterations $I_{\max}$, $A_m^{(0)}$, and $v_m^{(0)}$. 
  \Repeat
    \State Calculate $R(\bar{\mathbf{z}}^{(\rm{i})})$ and $\nabla_{\bar{\mathbf{z}}} R(\bar{\mathbf{z}}^{(\rm{i})})$;
    \State Solve convex subproblem (P2):
\begin{subequations}
\begin{alignat}{2}
&&&\text{(P2): }\underset{\{A_m,v_m\}_{m\in\mathcal{M}}}{\max} R(\bar{\mathbf{z}}^{(\rm{i})}) + \nabla_{\bar{\mathbf{z}}} R(\bar{\mathbf{z}}^{(\rm{i})})^T (\bar{\mathbf{z}} - \bar{\mathbf{z}}^{(\rm{i})})\nonumber \\[0.75mm]
&&&\text{s.t.} \ \ \ \ 0 \leq A_m \leq A_{\rm{max}}, \quad \forall m\in\mathcal{M},\nonumber \\[0.75mm]
&&& \ \ \ \ \ \ \ 0 \leq v_m \leq v_{\rm{max}}, \quad \forall m\in\mathcal{M},\nonumber \\[0.75mm]
&&& \ \ \ \ \ \ \ \left| A_m v_m^{(\rm{i})} + A_m^{(\rm{i})} v_m - A_m^{(\rm{i})} v_m^{(\rm{i})} \right| \leq 1, \quad \forall m\in\mathcal{M}.\nonumber
\end{alignat}
\end{subequations}
    \State Obtain $A_{m}^{(\rm{i}+1)}$ and $v_{m}^{(\rm{i}+1)}, \ \forall m \in \mathcal{M};$
    \State Update $\rm{i}= \rm{i} +1;$ 
  \Until
  	\State Convergence or maximum number of iterations $I_{\max}$ is reached.
\end{algorithmic}
\caption{Proposed Algorithm for the SRA array}
\end{algorithm}

\vspace{-4mm}
\section{Numerical Results}
\vspace{-1mm}
In this section, we evaluate via numerical simulations the proposed SRA array in terms of the achievable sum rate. To verify the performance, we consider the following benchmarks: (1) a fixed concentric circular antenna array (CCAA) realized by having the soft robot without deformation (refer to Fig. \ref{fig:noDeformation}), (2) a reconfigurable CCAA realized by antenna element repositioning along the $x$ and $y$ planes, and (3) a conceptual 3D CCAA that allows individual element movement along the $z$ plane; included purely as an upper bound benchmark, since no practical implementation of this challenging design has been proposed, contrary to the SRA whose physical realization has been validated~\cite{TipSR}. Users and scatterers are uniformly distributed with elevation and azimuth angles $\vartheta_{c, p, k} \in [0,\pi/2]$ and $\varphi_{c, p, k} \in [0, 2\pi]$. Unless stated otherwise, we set $M\!=\!8$, $N\!=\!4$, $L_{\rm{max}}\!=\!N\lambda/2$, $A_{\rm{max}}\!=\!0.2$, $v_{\rm{max}}\!=\!5 \ \text{m}^{-1}$, $I_{\rm{max}}\!=\!20$, $\varepsilon\!=\!10^{-4}$, $N_c\!=\!3$, $N_p\!=\!10$, and $\kappa=2$. We consider an operating frequency of $f_c=1.2$ GHz~\cite{TipSR}. The results are averaged over 1000 independent channel realizations.

The performance of the proposed array is first evaluated in Fig. \ref{fig:2a} in terms of sum rate with a signal-to-noise ratio (SNR) ranging from 0 to 30 dB. We notice that, for $K=2$, the proposed array achieves a 61.9$\%$ gain over the fixed CCAA and an 18.1$\%$ gain over the 2D reconfigurable CCAA at an $\text{SNR}=20$ dB. This notable performance gain is mainly attributed to the extra DoFs offered by the array structure flexibility. As expected, the 3D CCAA outperforms the proposed SRA by $11.1\%$ since the former enjoys more DoFs by moving each antenna element at the expense of much higher control complexity. In addition, the advantage of structure reconfigurability is more pronounced for $K=4$, as the SRA array adds around 12 bps/Hz over its fixed counterpart, compared to 7 bps/Hz for $K=2$. This result highlights that as multiuser interference becomes more challenging, having fine-grained control over the entire array achieves better performance.

The sum rate of the SRA array is evaluated in Fig. \ref{fig:2b} for different numbers of users $K$. As a general observation, all schemes exhibit a similar trend in sum rate as $K$ approaches the array's spatial DoFs (moving from an interference-limited to a power-limited regime). Nevertheless, we remark that the SRA outperforms the fixed and the 2D CCAA across all user loads by exploiting structure deformation to form sharper beams. Although the 3D CCAA achieves the highest peak sum rates, it requires more control variables, highlighting a performance-complexity tradeoff.

In Fig. \ref{fig:2c}, the sum rate is evaluated for different directivity factors, with an SNR ranging from 0 to 21 dB, for $K=4$. The proposed SRA array outperforms both the fixed CCAA and the 2D reconfigurable CCAA across the entire SNR range, though the 3D reconfigurable CCAA still achieves the highest rates. Furthermore, all antenna architectures experience an increase in sum rate for higher directivity values. However, we remark that the SRA array exhibits a $13.48\%$ improvement in sum rate, compared to an $11.34\%$ increase for the 2D CCAA, thereby highlighting the benefits of antenna structure reconfigurability over antenna element repositioning.

\section{Conclusion}
In this letter, we proposed a novel soft robot-inspired antenna array by considering tentacle-like structures with multiple antenna elements. Reconfigurability was achieved through geometry deformation by considering an amplitude and a spatial frequency parameter. A multi-user MISO downlink system was studied, in which the SCA was adopted to solve for the optimal deformation parameters that maximize the sum rate. Results showcased that the SRA array outperformed both fixed and per-element RA geometries by $26\%$ and $73\%$, respectively, unlocking a new degree of freedom through structure-level flexibility. Future works include generalizing to more complex network topologies and advanced deformation models that fully exploit structural flexibility.

\appendices
\section{Calculation of The Gradient $\nabla_{\bar{\mathbf{z}}} R(\bar{\mathbf{z}})$}
Based on the sum rate expression, $\nabla_{\mathbf{z}} R(\mathbf{z})$ is expressed as
\begin{equation}
\nabla_{\mathbf{z}} R(\mathbf{z}) = \sum_{k\in \mathcal{K}} \frac{1}{\ln(2)(1+\gamma_{k}(\mathbf{z}))} \nabla_{\mathbf{z}} \gamma_{k}(\mathbf{z}),
\vspace{-2mm}
\end{equation}
which is decomposed by applying the chain rule
\begin{align}
\frac{\partial R}{\partial z_{m,n}}
=\sum_{k\in \mathcal{K}} \frac{1}{\ln(2)(1+\gamma_k(z_{m,n}))} \sum_{i\in\mathcal{K}} \frac{\partial\gamma_k}{\partial S_{ki}}
\frac{\partial S_{ki}}{\partial z_{m,n}},
\end{align}
where $S_{ki}\triangleq\left|\mathbf{h}_k^H \mathbf{w}_i\right|^2$,
\begin{align}
\frac{\partial\gamma_k}{\partial S_{ki}}=
\begin{cases}
1/ (\sum_{i\in\mathcal{K}}S_{ki}+\sigma^2), & i=k,\\
-S_{kk}/(\sum_{i\in\mathcal{K}}S_{ki}+\sigma^2)^2, & i\neq k,
\end{cases}
\end{align}
and $\frac{\partial S_{ki}}{\partial z_{m,n}}
=2\Re\left\{(\mathbf{h}_k^H \mathbf{w}_i)\mathbf{w}_i^H \frac{\partial \mathbf{h}_k}{\partial z_{m,n}}\right\}$.
Moreover, each entry in $\mathbf{h}_k$ only depends on $z_{m,n}$ through $e^{-j\frac{2\pi}{\lambda}z_{m,n}\cos\vartheta}$. Thus, $\partial \mathbf{h}_k/\partial z_{m,n} = -j\frac{2\pi}{\lambda}\cos\vartheta \mathbf{h}_k$. Since $z_{m,n}(A_m,v_m)
= A_m\sin(\omega t + v_m\ell_n)$, the corresponding partial derivatives are
\begin{align}
\frac{\partial z_{m,n}}{\partial A_m}
=\sin(\omega t + v_m\ell_n), \ \frac{\partial z_{m,n}}{\partial v_m}
= A_m\ell_n\cos(\omega t + v_m\ell_n).
\vspace{-1mm}
\end{align}
By defining the intermediate gradient as $\bar{\mathbf{g}} = \bigl[\frac{\partial R}{\partial z_{1,1}},\dots,\frac{\partial R}{\partial z_{M,N}}\bigr]^T \in \mathbb{R}^{MN\times 1}$,
and the Jacobian matrix $\mathbf{J}_m \in \mathbb{R}^{N\times 2}$ for the $m$-th tentacle as
\begin{align}
\mathbf{J}_m =
\begin{bmatrix}
\sin(\omega t+v_m\ell_1), & A_m\ell_1\cos(\omega t+v_m\ell_1)\\[1mm]
\vdots & \vdots\\[1mm]
\sin(\omega t+v_m\ell_N), & A_m\ell_N\cos(\omega t+v_m\ell_N)
\end{bmatrix},
\end{align}
we obtain $\nabla_{\bar{\mathbf{z}}} R(\bar{\mathbf{z}})
= \mathbf{J}^T\,\bar{\mathbf{g}} \ \in \mathbb{R}^{2M\times 1}$, where $\mathbf{J} = \mathrm{diag}(\mathbf{J}_1,\dots,\mathbf{J}_M)\in\mathbb R^{MN\times2M}$.


\begin{thebibliography}{00}
\bibitem{6G} H. Tataria, M. Shafi, A. F. Molisch, M. Dohler, H. Sj\"{o}land, and F. Tufvesson, ``6G wireless systems: Vision, requirements, challenges, insights, and opportunities,” \emph{Proc. IEEE}, vol. 109, no. 7, pp. 1166–1199, Jul. 2021.

\bibitem{RIS} H. Hui, Y. Zou, Y. Li, L. Zhai, and B. Ning, ``Robust beamforming design for RIS-assisted cognitive radio systems with hardware impairments," \emph{IEEE Trans. Veh. Technol.}, vol. 73, no. 12, pp. 19080-19095, Dec. 2024.

\bibitem{FA} K.-K. Wong, A. Shojaeifard, K.-F. Tong, and Y. Zhang, ``Fluid antenna systems,” \emph{IEEE Trans. Wireless Commun.}, vol. 20, no. 3, pp. 1950–1962, Mar. 2021.

\bibitem{MA} L. Zhu, W. Ma, and R. Zhang, ``Modeling and performance analysis for movable antenna enabled wireless communications,” \emph{IEEE Trans. Wireless Commun.}, vol. 23, no. 6, pp. 6234–6250, Jun. 2024.

\bibitem{FlAA} S. Yang, J. An, Y. Xiu, W. Lyu, B. Ning, Z. Zhang, M. Debbah, and C. Yuen, ``Flexible antenna arrays for wireless communications: Modeling and performance evaluation,” \emph{IEEE Trans. Wireless Commun.}, vol. 24, no. 6, pp. 4937-4951, Jun. 2025.

\bibitem{FIM} J. An, C. Yuen, M. D. Renzo, M. Debbah, H. Vincent Poor, and L. Hanzo, ``Flexible intelligent metasurfaces for downlink multiuser MISO communications,” \emph{IEEE Trans. Wireless Commun.}, vol. 24, no. 4, pp. 2940-2955, Apr. 2025.

\bibitem{FAtut} W. K. New et al., ``A tutorial on fluid antenna system for 6G networks: Encompassing communication theory, optimization methods and hardware designs,” \emph{IEEE Commun. Surv. Tuts.}, early access, Nov. 15, 2024.

\bibitem{MAtut} L. Zhu, W. Ma and R. Zhang, ``Movable antennas for wireless communication: Opportunities and challenges", \emph{IEEE Commun. Mag.}, vol. 62, no. 6, pp. 114-120, Jun. 2024.

\bibitem{FAMIMO} C. N. Efrem and I. Krikidis, ``Transmit and receive antenna port selection for channel capacity maximization in fluid-MIMO systems", \emph{IEEE Wireless Commun. Lett.}, vol. 13, no. 11, pp. 3202–3206, Nov. 2024.

\bibitem{PowConMA} Y. Wu, D. Xu, D. W. K. Ng, W. Gerstacker, and R. Schober, ``Globally optimal movable antenna-enhanced multi-user communication: Discrete antenna positioning, motion power consumption, and imperfect CSI,” \emph{arXiv preprint arXiv:2408.15435}, Aug. 2024.

\bibitem{FIMSensing} Z. Teng, J. An, L. Gan, N. Al-Dhahir, and Z. Han, ``Flexible intelligent metasurface for enhancing multi-target wireless sensing", \emph{IEEE Trans. Veh. Technol.}, early access, Jul. 2025.

\bibitem{SRnature} D. Rus and M. T. Tolley, ``Design, fabrication and control of soft robots,” \emph{Nature}, vol. 521, no. 7553, pp. 467–475, May 2015.

\bibitem{TipSR} L. H. Blumenschein, L. T. Gan, J. A. Fan, A. M. Okamura, and E. W. Hawkes, ``A tip-extending soft robot enables reconfigurable and deployable antennas,” \emph{IEEE Robot. Automat. Lett.}, vol. 3, no. 2, pp. 949–956, Apr. 2018.

\bibitem{RotSR} F. Deng, S. Yao and K. M. Luk, ``Rotenna: Harnessing seamless integration of continuum robot for dynamic electromagnetic reconfiguration", \emph{IEEE Robot. Automat. Lett.}, vol. 8, no. 12, pp. 8557-8564, Dec. 2023.

\bibitem{BalAnt} C. A. Balanis, \emph{Antenna theory: Analysis and design}, 2nd ed. John
Wiley \& Sons, Inc., 1997.
\end{thebibliography}
\end{document}